# Online shopping key features analysis in Mures county

Author(s)*: Elena-Iulia APĂVĂLOAIE [1], Liviu Onoriu MARIAN [2], Elena Lucia HARPA [3]
Position: PhD Student [1,3], Prof., PhD [2]
University: Technical University of Cluj-Napoca [1, 3], "Petru Maior" University of Tîrgu Mureș [2]
Address: Cluj-Napoca, Memorandumului Str., No. 28, Romania[1,3]
Târgu Mureş, Nicolae Iorga Str., No. 1, Romania[2]
Email: iulia8@gmail.com [1], liviu.marian@yahoo.com [2], elena_harpa@yahoo.com [3]
Webpage: http://www.utcluj.ro/ ; http://upm.ro/

**Abstract**

*Purpose* – The aim of this paper is to get an overview of the online buyer profile, and also some key aspects in the way the online shopping is conducted.
*Methodology/approach* – In this project we conducted a quantitative research, consisting of a questionnaire based survey. For data processing and interpretation we used SPSS statistical software and Excel. For data analysis, we used the descriptive statistics indicators, and a series of bi-varied analysis for testing some statistical assumptions.
*Findings* – Viewed at first with skepticism by the Internet users in Romania, because of the many news about how dangerous the credit card payments are, the online stores have gained much ground and trust in the recent years.
*Research limitations/implications* – Since the study was conducted mainly in the online environment, we can not talk about the representativeness of the sample, only about a trend observed in the studied population.
*Practical implications* – The study helps us understand the population reactions and attitudes regarding the online shopping.
*Originality/value* – The study revealed some important issues regarding the online shopping in Mures county, issues that are described in detail in the content of this paper.
*Key words:* e-commerce, online shopping, impact of the Internet.

## Introduction

The penetration of the Internet in everyday life has created a new type of economy (net-economy) where the technology connects anyone to anything and the dominating features are communications, standards and open markets (Grosseck, 2006).

Electronic commerce (e-commerce) has quickly become a strategally important tool for businesses. The e-commerce provides to the seller a less expensive means of growth and to the customer easier access to a wide range of products and services (Florescu, 2007).

There are many definitions of e-commerce, but broadly speaking, it is a concept that indicates the purchase and sale or exchange of goods, services, information, through the Internet.

A concept that is increasingly widely used in the Internet commerce is mobile commerce. Less well known is the concept of social commerce.

The number of people who adopt this type of shopping it's increasing, and the good news is that on the Internet we can find almost the same wide range of products / services as in the traditional commerce.

According to the official report issued by GPeC[1], the value of the products sold online in Romania in 2013 was approximately 600 million euros, a figure similar to the one recorded in 2012. This figure does not include services, utilities payment, airline tickets or tourist services. Although most people prefer to pay cash (this method is used in 90% of the transactions), there is an increase of 35% compared to 2012 of the transactions paid by card (figure 1).

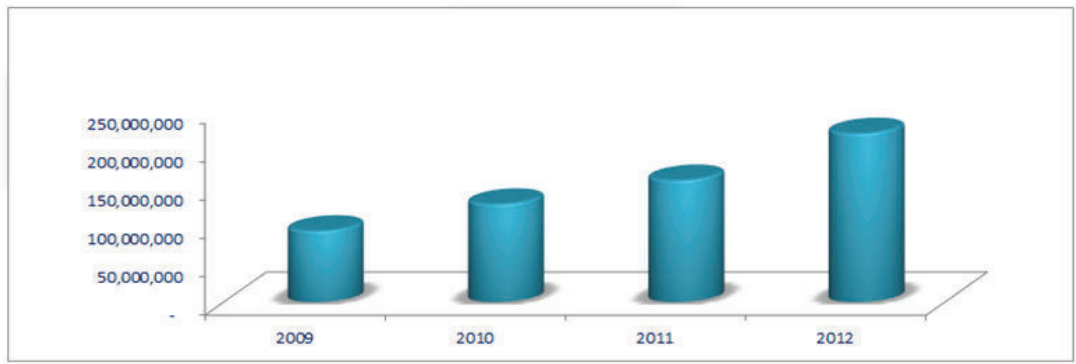

Figure 1 – The evolution of the credit card payments in Romania (euro)
Source: www.payu.ro

In an era where connecting to information has become a necessity, the electronic commerce is a challenge. The development of information and communication technologies enables commercial transactions in a very short time and without intermediaries. The electronic commerce allows the consumers free access to information, and they can become the owners of the goods in fractions of a second.

## Methodology

The research involved going through five **stages** (Kotler, 2008), as follows:
1. **Defining the problem and the research objectives** - as I mentioned before, this paper contains a questionnaire based survey regarding to online shopping in Mures County.

   The main **objectives** of the study were:
   - Identifying the amount of Internet users that are conducting online shopping;
   - Identifying the main characteristics of online purchasing behavior (whether they are buying, how often, what are they buying, how much they spend, payment methods, delivery methods);
   - Identifying the main reasons for preferring online shopping rather than traditional shopping;
   - Identifying the main factors which influence the online purchase decision;
   - Identifying the satisfaction level of the transactions made in the virtual environment;
   - Identifying general attitudes and perceptions regarding online shopping (respondents trust level regarding the online environment, especially the electronic payments, analyze the implications of certain statements related to online shopping and consumer decision);
   - Identifying the main problems / disadvantages encountered in the online shopping;
   - Highlighting the consumer perspective related to the e-commerce future;
   - Identification of the online buyer profile.

   Based on these objectives, **the hypotheses** of the study were:
   - The respondents use the Internet for online shopping;
   - The respondents take advantages of the fact that online stores have flexible hours (24 hours of 24, 7 days out of 7);
   - The online buyers are guided by the lowest prices when it comes to both products and transportation;

---

[1] GPeC (Gala Premiilor e-Commerce), is one of the most important event of e-commerce in Romania

- The main categories of products purchased through the Internet are Electronics / IT, Flights / hotel reservations and Clothing;
- The majority of the online consumers prefer to pay when they receive the goods they ordered;
- The majority of the respondents considered online payments to be as safe as the traditional payments;
- The inability to touch / try the products is the main disadvantage of the online shopping, while saving time is the main reason why respondents choose to buy on the Internet;
- Most of the online buyers are young, educated and living in the urban areas.

2. **Development of the research plan** - this stage involved making decisions about data sources, research methods, research tools, sampling and methods of contacting respondents.

   So, for primary data collection I used a questionnaire that was pre-tested on a sample of subjects. In the survey, questions were formulated according to the objectives and purposes of the research and more types of questions were used: factual questions, closed questions with only one answer, closed questions with multiple answers, test questions, etc (Ciucan-Rusu, et all, 2011).

   The questionnaire includes questions with more types of scales, like: *Likert scale*, *Stapel scale, importance scale, appreciation scale, etc.* Among the variables related to socio-demographic characteristics of the sample we used the following: the gender of respondent, age, nationality, marital status, place of origin, studies, employment status, monthly income of the respondent.

3. **Data collection** – was performed both virtual path and directly, in printed version, the distribution of the respondents being influenced by their willingness to complete the questionnaire. The electronic questionnaire was conducted using a custom software based on WordPress, and the answers of the respondents were saved in a database, which was after that imported into Excel and SPSS. The sample included 271 people, both urban and rural, potential online buyers, and data collection was conducted over a period of 14 days.

4. **Data analysis** – Before inserting the data into SPSS, I checked the questionnaire filling level and eliminated the ones which were incomplete. For the data processing stage I used SPSS software, which is one of the most popular solutions for data analysis (Stavarache, 2005). For interpreting the results I used Microsoft Excel. For data analysis, we used the descriptive statistics indicators (absolute frequencies, relative frequencies, simple arithmetic mean, and weighted arithmetic mean), and a series of bi-varied analysis for testing some statistical assumptions (chi square test, Kendall and Spearman correlation coefficients, binomial test and principal component analysis).

5. **Findings/Conclusions** – the study led to some concrete results regarding the online shopping in Mures county, results that completes the picture of the county on using the Internet and which will be presented in what follows.

### The presentation of the sample used in the research

The sample included a number of 271 respondents, with the following structure (figure 2) in terms of their place of origin, the aim being to have a proportional distribution of the questionnaires with the actual situation of the online buyers, namely major from urban areas.

As it can be seen in the figure above, 86,4% of the respondents come from urban areas, while only 13,6% of them come from rural areas.

Further I will present the structure of the sample according to the socio - demographic characteristics.

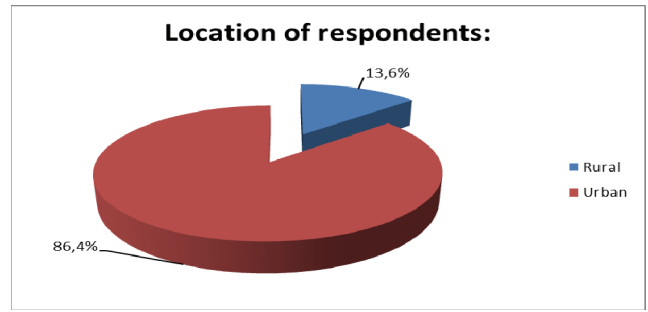

Figure 2 – Place of origin

The structure of the sample by gender is almost equal, i.e. 52,3% female and 47,7% male, according to figure 3. Regarding the marital status of respondents, most of them are unmarried, i.e. 55,1%, followed by those who are married (37,2%). The least of them are living in free union or are divorced or widowed, according to figure 4.

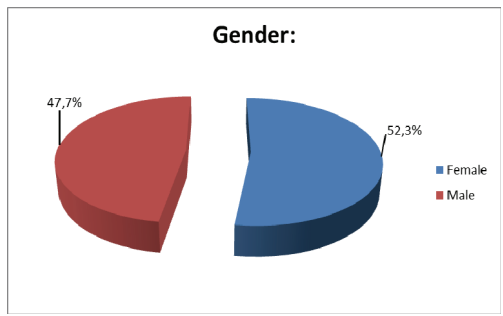
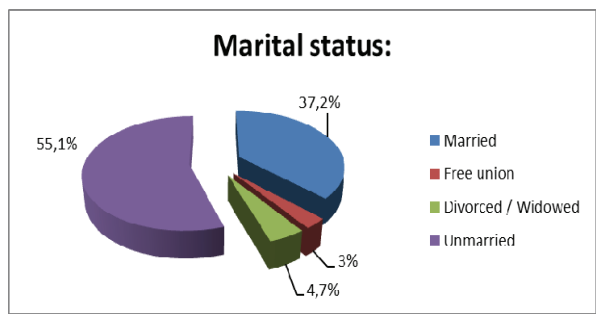

Figure 3 – Gender                   Figure 4 – Marital status

In the sample of respondents are represented all categories of occupations identified as initial target, mostly employees with higher education (60,3%), followed by students (20,5%), managers (6,4%) and employees with average education (6%), as shown in figure 5.

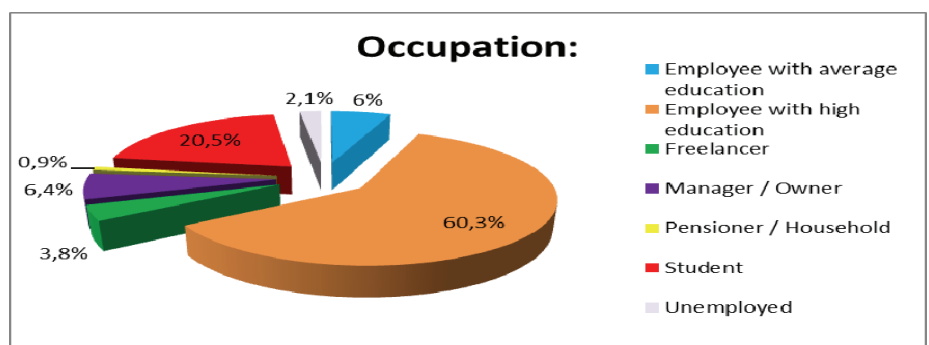

Figure 5 - Occupation

Closely related to the occupation of the respondents is their income level, depending on which the sample has the following structure. (Figure 6)

As it can be observed in figure 6, approximately one third of the respondents (30.3%) have an income of over 2.500 lei, followed by those with no income (13.9%) and those who earn monthly between 1101 and 1400 lei (11.7%). This situation is correlated with the fact that most of the respondents are employed with higher education, which means higher incomes, followed by students who do not have yet a fixed job.

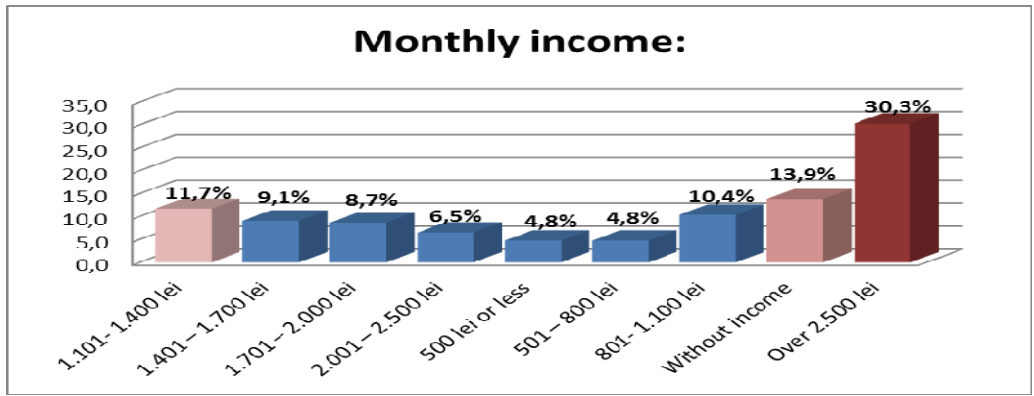

Figure 6 – Income

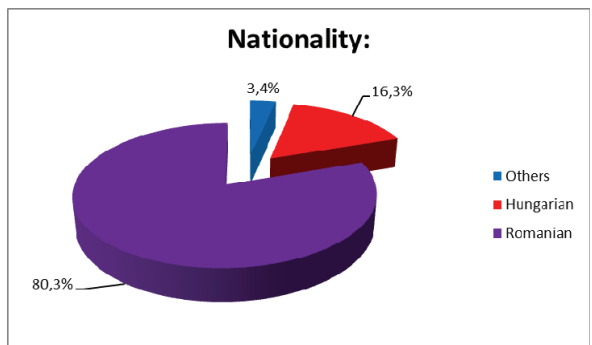

Figure 7 – Nationality

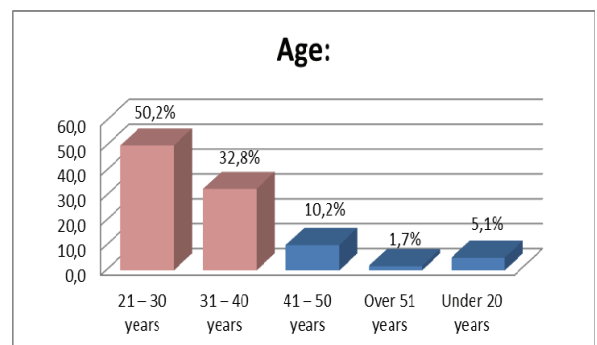

Figure 8 - Age

The sample structure according to nationality, groups the respondents as follows: 80,3% Romanian users, 16,3% Hungarian users and 3,4% other nationalities (i.e. German, Moldavian and Roma), as shown in figure 7.

Regarding the sample structure by age, the majority of respondents were aged between 21 and 30 years (43.5%), followed by those aged between 31 and 40 years (24.8%), those under 20 years (14.6%), those between 41 and 50 years (12.9%) and the least (4.2%) were those aged over 51 years (figure 7).

### Findings of the study

The first results presented refer to the percentage of online buyers among the respondents. As shown in figure 9, 88,2% of survey participants are online buyers, while only 11,8% of them don't use the Internet for shopping.

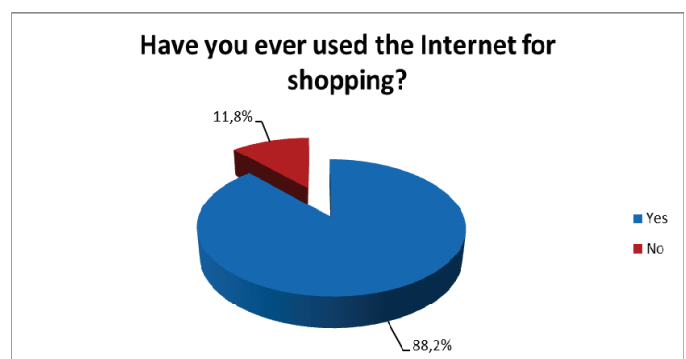

Figure 9 - Percentage of the online buyers

Regarding the frequency of the online shopping (Figure 10), the majority of the respondents (23,8%) shop online every six months, followed by those who shop online every three months (22,2%) and once a month (3,3%). Only 1,7% of respondents have not purchased anything in the last 12 months.

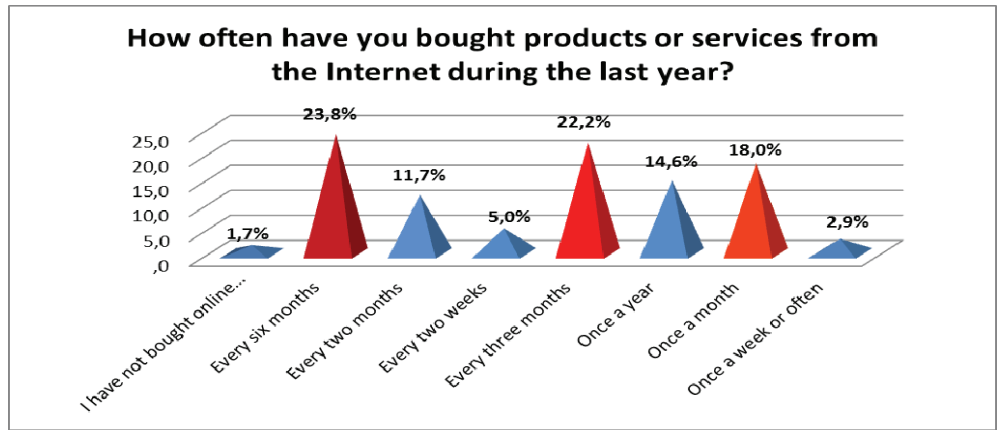

Figure 10 – Frequency of the online buying

Another important aspect related to the frequency of online buying is the time interval on which the respondents make these transactions.

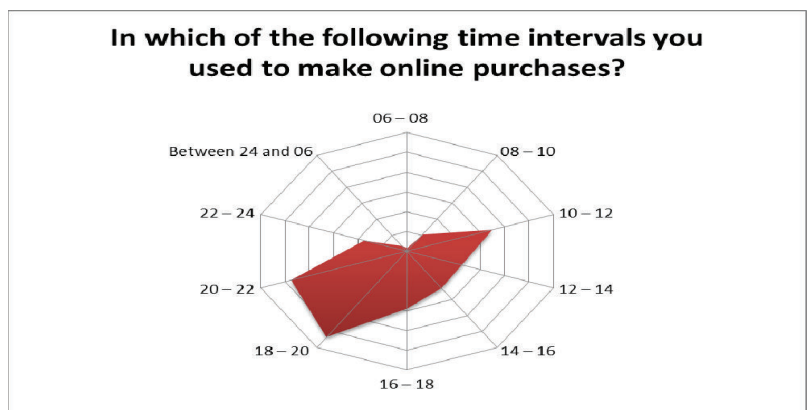

Figure 11 – Time intervals

Despite the idea that on the Internet you can buy at any time of day or night, because the online stores are open 24 hours of 24, this research shows that most of the respondents are shopping online between 18-20 (107 persons) and between 20-22 (95 people). None of the respondents shopped between 24-08 (see figure 11).

Regarding the reason why the respondents chose to buy online, the situation is as follows in Figure 12.

At small distances between them, the main reasons why the respondents chose to buy online instead of the traditional stores are: availability of cheaper products online (53,6%), saving time (52,7%) and access to certain products which are not available in the traditional stores (52,3%). Other reasons include the possibility to order at any hour or day of the week (48,5%) and price comparisons (37,2%). (Figure 13)

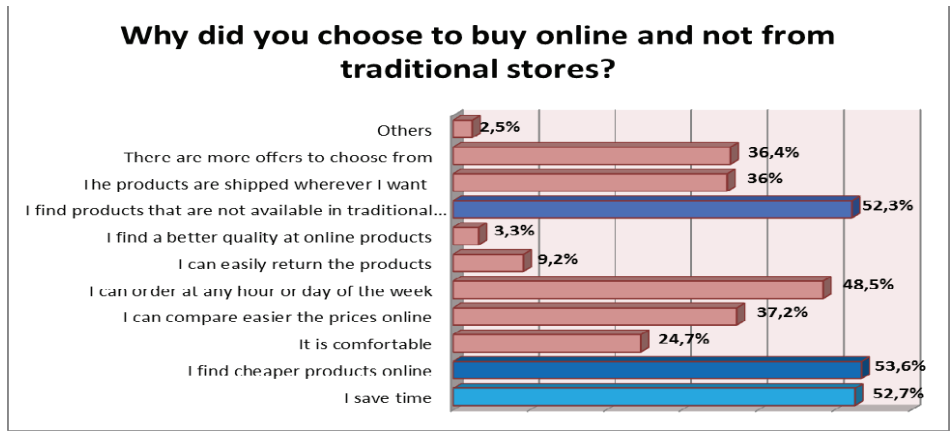

Figure 12 – The reason of choosing online shopping

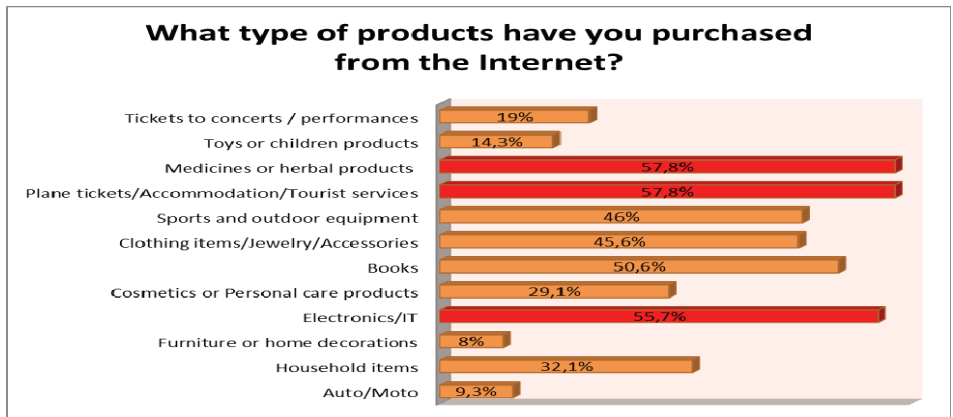

Figure 13 – Types of products bought online

One of the objectives of the study was to identify the main categories of products purchased on the Internet. The analysis of the answers revealed that more than half of the respondents have purchased Plane tickets/Accommodation/Tourist services (57,8%), Medicines or natural products (57,8%) and Electronics/IT (55,7%). Fewest bought Furniture or home decorations (8%) and Auto/Moto (9,3%).

The last result presented in this paper refers to the importance of certain criteria in making purchasing decision.

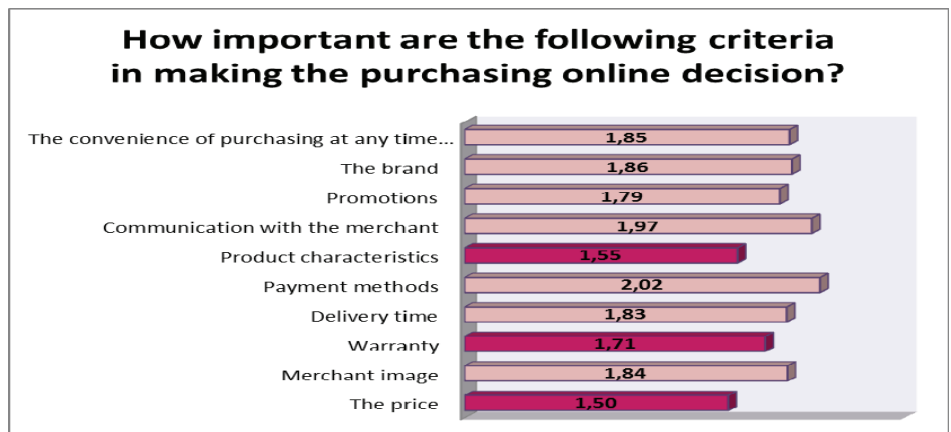

Figure 14 – The importance of certain criteria

According to figure 14, the most important criteria in the decision making is the price, followed by product characteristics, warranty, promotions and delivery time.

## Conclusions

The online buyer profile, according to the survey looks like this: woman, unmarried, from urban area, aged between 21 and 30 years, employee with higher education.

Most of the respondents choose to buy online for the following reasons: the existence of cheaper products in the online environment (53.6%), saving time (52.7%) and access to certain products that do not exist in physical stores (52.3 %).

The most important criteria in the decision of buying online is the price, followed by product features and warranty.

Almost half of the respondents (49.8%) are satisfied with the online transactions they made, while the other half (48%) are very satisfied.

In this study we started from several assumptions and after analyzing the data four of them were right, two of them were wrong, and two of them were partial confirmed.

The e-commerce development is held back by certain aspects such as the population lack of computerization (especially in the rural areas) and communications infrastructure which is not developed enough. Even so, in the recent years, the number of virtual stores in Romania has increased due to the promotional offers and the convenience of the people.